\documentclass[a4paper,accepted=2024-04-11]{quantumarticle}
\pdfoutput=1

\usepackage[utf8]{inputenc}
\usepackage[T1]{fontenc}
\usepackage[compress]{cite}
\usepackage{array}
\usepackage{amsmath}
\usepackage{color,soul}
\PassOptionsToPackage{hyphens}{url}
\usepackage{hyperref}
\usepackage{url}
\usepackage{dsfont}
\usepackage{subfig}
\usepackage{graphicx}
\usepackage[ruled]{algorithm2e}
\usepackage[dvipsnames]{xcolor}
\usepackage{float}
\usepackage{soul}
\usepackage{eufrak}

\SetAlFnt{\small}
\SetAlCapFnt{\small}
\SetAlCapNameFnt{\small}
\SetKwInput{KwInput}{Input}
\SetKwInput{KwOutput}{Output}

\DeclareMathOperator*{\minsec}{\ensuremath{min_2}}
\newcommand{\be}{\mathbf{e}}
\newcommand{\bs}{\mathbf{s}}
\newcommand{\bH}{\mathbf{H}}
\newcommand{\Nbh}{\mathcal{N}}
\newcommand{\bRel}{\mathbf{\gamma}}
\newcommand{\etak}{\mathfrak{N}_k}

\newcommand{\V}{Q} 
\renewcommand{\v}{q} 

\graphicspath{{data/}}

\begin{document}

\title{Check-Agnosia based Post-Processor for Message-Passing Decoding of Quantum LDPC Codes}

\author{Julien du Crest}
\affiliation{Universit\'e Grenoble Alpes, Grenoble INP, LIG, F-38000 Grenoble, France}
\email{julien.du-crest@univ-grenoble-alpes.fr}
\author{Francisco Garcia-Herrero}
\affiliation{Department of Computer Architecture and Automatics, Complutense University of Madrid, Madrid, Spain}
\email{francg18@ucm.es}
\author{Mehdi Mhalla}
\affiliation{Universit\'e Grenoble Alpes, CNRS, Grenoble INP, LIG, F-38000 Grenoble, France}
\email{mehdi.mhalla@univ-grenoble-alpes.fr}
\author{Valentin Savin}
\affiliation{Universit\'e Grenoble Alpes, CEA-L\'eti, F-38054 Grenoble, France}
\email{valentin.savin@cea.fr}
\author{Javier Valls}
\affiliation{Instituto de Telecomunicaciones y Aplicaciones Multimedia, Universitat Politecnica de Valencia, Valencia, Spain}
\email{jvalls@upv.es}

\begin{abstract}
The inherent degeneracy of quantum low-density parity-check codes poses a challenge to their decoding,  as it significantly degrades the error-correction performance of classical message-passing decoders. To improve 
their performance, 
a post-processing algorithm is usually employed.
To narrow the gap between algorithmic solutions and hardware limitations,
we introduce a new post-processing algorithm with a hardware-friendly orientation, providing error correction performance competitive to the state-of-the-art techniques.  The proposed post-processing, referred to as check-agnosia, is inspired by stabilizer-inactivation, while considerably reducing the required hardware resources, and providing enough flexibility to allow different message-passing schedules and hardware architectures.
We carry out a detailed analysis for a set of Pareto architectures with different tradeoffs between latency and power consumption, derived from the results of implemented designs on an FPGA board. We show that latency values close to one microsecond can be obtained on the FPGA board, and provide evidence that much lower latency values can be obtained for ASIC implementations. 
In the process, we also demonstrate the practical implications of the recently introduced t-covering layers and random-order layered scheduling. 
\end{abstract}

\maketitle


\section{Introduction}\label{sec:intro}

Quantum low-density parity-check (qLDPC) codes \cite{15yearsQLDPC} have become one of the main candidates to implement the error-correction layer of a large-scale quantum computer architecture~\cite{gottesman2014fault, ChallengesZooImplementQLDPC, bravyi2022future}. 
Compared to other families of quantum error correction codes, qLDPC codes may reduce the physical qubit overhead, while protecting a larger number of logical qubits, so higher code rates can be obtained with similar or better error-correction performance \cite{DegenerateQLDPC, BalancedQLDPC, leverrier2022quantum, RateDistQLDPC, bravyi2023high}. 
Yet, for qLDPC codes to work on a real system, a larger number of physical qubits than those available in today's noisy intermediate-scale quantum systems is required~\cite{ChallengesZooImplementQLDPC}, \cite{TowardsQLDPC}. 
Nonetheless, before large-scale quantum technology becomes available, two important problems need to be addressed from the qLDPC decoding perspective: i) devising new decoding algorithms that overcome or mitigate the effect of degeneracy~\cite{QuantumTrappingSets}, thus providing increased error correction capabilities, and ii) developing hardware designs that meet latency and power constraints imposed by the quantum system (\emph{e.g.}, latency values within the decoherence time of the qubits to be protected, or power limitations for qubit technologies requiring cryogenic cooling,  when the decoder is implemented within the low-temperature layers),  a topic that only got attention recently \cite{HWSurveyQEC, 23Turbo-XZ, Delfosse2021almostlineartime}.
%



To achieve the first objective, several approaches building upon classical message-passing (MP) decoding algorithms have been recently proposed in the literature,  where the degeneracy issue is dealt with by either incorporating neural network techniques in the MP decoder \cite{NNBP_QEC}, or adding a post-processing step, taking advantage of the soft information delivered by the MP decoder \cite{DegenerateQLDPC, StabilizerInactivationQLDPC}.

Neural-network-based decoders are bound to the noise models used to train them and do not scale well with the number of qubits \cite{NeuralFPGAScale}. Moreover, as shown in \cite{NeuralNoiseExp}, there are not only different sources and noise models, but also the noise may be different depending on the area of the layout of the quantum processor, the environmental conditions, and the evolution of errors with time since the last calibration (space and time drift of the errors \cite{NoiseDrift}). In that sense, more generalized solutions are required, at least at the moment of writing these lines, when there is no standardized or predominant technology or architecture for future large-scale quantum devices. 
Hence, post-processing techniques may become an interesting choice. 
A first post-processing technique based on ordered statistics decoding (OSD) was proposed in \cite{DegenerateQLDPC, BPOSD}. Although the improvement in terms of coding gain is significant, the complexity is too high, and hence, it becomes unpractical for real-time hardware implementations \cite{OSD_FPGA}. Recently, some of us proposed a new post-processing technique for Calderbank-Shor-Steane (CSS) qLDPC codes, called stabilizer inactivation \cite{StabilizerInactivationQLDPC}.
The post-processing consists in \emph{inactivating} a set of unreliable qubits supporting a check in the dual code (a stabilizer generator of the same type as the decoded error). Then the MP decoding is run again, while taking out of the decoding process the inactivated qubits and their neighbor check nodes. The remaining qubits and check nodes that participate in the MP decoding are called active. Several stabilizer generators may be inactivated, one at a time (which can be implemented either sequentially or in parallel), until one MP decoding meets the syndrome constraints on the active check nodes. Inactivated qubits are then determined by solving a small linear system, defined by the inactive check-nodes.  Stabilizer inactivation shows a non-negligible error correction improvement and increased flexibility (with regard to the MP decoding schedule) compared to OSD, with a considerable reduction of complexity. 
However, as discussed later in this paper, 
additional hardware-oriented analysis and optimization are required to ensure the hardware design meets the constraints required to provide real-time support to a quantum processor.

\medskip The main contributions of the paper are as follows.
First, inspired by the stabilizer inactivation, we introduce a new post-processing algorithm for MP decoders. The algorithm takes into account the architectural properties of MP decoders in order to reduce the computational load and the required hardware resources. It also limits the amount of information required from the code, eliminating the need to know the stabilizer structure (dual code) and just treating both parity-check matrices as independent. 
Similar to the stabilizer inactivation, the algorithm identifies a small set of unreliable qubits (which are however not inactivated, in the sense described above). The information considered to identify such a set of qubits is based on the check-node reliability. When the MP decoder fails, the a priori information for the qubits connected to the least reliable check nodes is erased, and the
%
post-processor will then try to learn again the reliability of these qubits based on the information from the rest. For this reason, we call the post-processing technique check-agnosia. We also suggest several approaches to perform the selection of unreliable check-nodes, to reduce power consumption and latency, which are the constraints that limit the implementation of decoders in real systems \cite{PowerLimit, TimingLimit}.  

Second, along the document, \textcolor{black}{a non-agnostic hardware perspective is described to help to meet the constraints of future large-scale quantum devices}. Aligned with this, a functional description in terms of performance and hardware results of the proposed solution for the two main schedules employed for MP decoders (flooded and layered~\cite{EfficientMPSchedule}, \cite{VLSISchedules}) is introduced.  We carry out a detailed analysis of different corner cases, which is then illustrated for a specific qLPDC code, by providing latency and power consumption values of the check-agnosia solution implemented on an FPGA board.



\medskip  The rest of the paper is organized as follows. Section~\ref{sec:algos} introduces the relevant notation and the algorithmic background. Section~\ref{sec:arch} introduces the check-agnosia post-processing, and discusses the check-node reliability metric along with several hardware-oriented optimizations.  
Section~\ref{sec:solution}  analyzes the impact of the post-processing algorithm on the hardware implementation, considering architectures with different schedules and varying degrees of parallelism. Latency and power consumption results are also provided here. Section \ref{sec:result} evaluates the error-correction performance of the proposed check-agnosia decoder for different qLDPC codes, and compares it to other existing solutions.  Finally, Section~\ref{sec:conclusion} provides the main conclusions of this work.

\section{Algorithmic Background}\label{sec:algos}

We consider qLDPC codes of CSS type, defined by two parity check matrices $\bH_x$ and $\bH_z$, corresponding respectively to $X$-type and $Z$-type generators. In the following, we will consider decoding of one type of error (since similar considerations apply to the other type), and will denote by $\bH$ the corresponding decoding matrix (\emph{e.g.}, $\bH=\bH_z$ for decoding $X$-type errors). We also consider the Tanner graph associated with $\bH$, and denote by  $\mathcal{\V}$  the set of qubit-nodes\footnote{Usually referred to as variable-nodes or bit-nodes, in classical LDPC coding.} and $\mathcal{C}$ is the set of check-nodes (stabilizer generators). We denote by $\mathcal{N}(\v) \subset \mathcal{C}$ the set of neighboring check-nodes of a qubit-node $\v \in \mathcal{\V}$, and similarly, by $\mathcal{N}(c) \subset \mathcal{\V}$ the set of neighboring qubit-nodes of a check-node $c\in  \mathcal{C}$. 


We denote by $p$ the probability that an error of the considered type occurs, and by $e\in \{0,1\}^{|\mathcal{\V}|}$ the error indicator vector\footnote{For a Pauli noise model in which Pauli errors $X, Y$, and $Z$ occur with probabilities $p_x, p_y$, and $p_z$, respectively, and considering the decoding of the $X$-type error, we have $p = p_x + p_y$, and $e_\v = 1$ iff either an $X$ or $Y$ error occurred on the corresponding qubit.}. We assume errors happen independently on the qubits, hence $P(e_{\v} = 1) = p$. Information about the error $\be$ is revealed through the measurement of stabilizer generators, in the form of an error syndrome $\bs := \bH \cdot \be$. Throughout this work, we assume ideal syndrome extraction, \textit{i.e.}, we only consider errors occurring on qubits, not on the extracted syndrome.  The decoding problem is determining the most likely error $\hat{\be}$, such that $\bH \cdot \hat{\be} = \bs$.  

Although maximum likelihood decoding is optimal, it is computationally prohibitive. Instead, classical LDPC codes are efficiently decoded by MP algorithms.
For a qubit ${\v}\in \mathcal{\V}$, we denote by $\bRel_{\v}$  the a priori log-likelihood ratio (LLR) of an error happening on qubit $\v$, which is defined\footnote{For a Pauli noise model, correlations between $X$ and $Z$ errors (due to the $Y$ errors) can be taken into account by decoding $X$ and  $Z$ errors sequentially, say first $X$ and then the $Z$ error, and computing the a priori LLRs for the $Z$ error, conditional on the decoded $X$ error.} as $\bRel_{\v} = \log\left(P(e_\v = 0) / P(e_\v = 1)\right) = \log((1-p)/p)$. The set of LLR values $\{\bRel_{\v} \mid \v\in \mathcal{\V}\}$ constitutes the input of the MP decoder and is used to initialize an iterative exchange of messages between qubit and check-nodes. We denote these messages by either  $\mu_{\v\rightarrow c}$ or $\mu_{c\rightarrow \v}$, the arrow in the notation indicating whether the message is sent from a qubit-node $\v$ to a check-node $c$, or in the opposite direction. At each iteration, exchanged messages are used to compute a posteriori LLR values $\hat{\bRel}_{\v}$, for each qubit-node $\v$, used to provide an estimate $\hat{\be}_\v$ of the corresponding qubit error. The iterative message passing process stops when either the estimated error satisfies the syndrome (\emph{i.e.}, $\bH\cdot \hat{\be} = \bs$), or a maximum number of iterations is reached. 
In the following, we shall also refer to (a priori/a posteriori) LLR values as \emph{qubit reliabilities}.

Throughout this work, we shall consider the Min-Sum (MS) decoding, or its normalized variant (NMS), which represent the option of choice for hardware implementations for several reasons: reduced computational complexity, reduced memory requirements (by adopting the first and second minimum compression method for check-node messages~\cite{boutillon2014hardware}), and its insensitivity to input LLR values up to a constant scaling factor (see Section~\ref{sec:result} for the input LLRs of the finite-precision MS decoder). For details regarding the MS and other MP decoding algorithms we refer to \cite{fbvalentin} (see also the discussion in \cite[Section 2]{StabilizerInactivationQLDPC}).

A key attribute of MP decoding algorithms is the underlying {\em scheduling}, indicating the order in which qubit - and check-node messages are updated~\cite{fbvalentin}. Flooded, layered, or serial schedules\footnote{Through this work, all schedules are considered to be horizontal, \emph{i.e.}, defined with respect to check-node processing, as opposed to vertical schedules, defined  with respect to qubit-node processing.} are usually implemented through
fully-parallel, partially-parallel, or serial hardware architectures, respectively, yielding designs with different performances in terms of latency, area, or power consumption. 

Compared to the flooded schedule, serial and layered schedules are also known to propagate information twice faster in the Tanner graph~\cite{zhang2007iterative} for classical (non-degenerate) LDPC codes. This directly translates into a faster convergence speed. However, the flooded schedule provides decoding performance similar to the serial and layered ones, at the cost of doubling the number of decoding iterations. 
This is no longer true for qLDPC codes, presumably due to the code degeneracy. As observed in~\cite{StabilizerInactivationQLDPC}, not only the flooded schedule may not be able to approach the decoding performance of the serial or layered schedules, even at the cost of an increased number of iterations, but in some cases it may also penalize the performance of the post-processing algorithm. Layered MP decoding of qLDPC codes has been recently investigated by the authors in~\cite{23Layered}, where it has also been observed that processing the layers in a random order (at each decoding iteration) may significantly improve the performance of the MP decoder. We will use these results in Section~\ref{sec:result} of this paper.

\section{Check-Agnosia Decoder}\label{sec:arch}
We introduce in this section the Check-Agnosia (CA) post-processing. 
We first describe the generic post-processing technique (Algorithm~1) and then discuss possible modifications.



\subsection{Generic Check-Agnosia Decoder}
The error vector $\hat{\mathbf{e}} $ is estimated first using a soft-output MP decoding algorithm. 
If the error estimate $\hat{\mathbf{e}} $ satisfies the syndrome, \emph{i.e.},  $\mathbf{H} \cdot \hat{\mathbf{e}}  =\bs$, then 
no post-processing is applied. 

If the initial MP decoding fails, a metric on the exchanged soft information is used to find the $\lambda$ checks $\{c_k\}_{k\in [\lambda]}$ whose supports are the most likely to be involved in the decoding failure (this metric will be discussed later). The post-processing will consist of rerunning the MP decoder at most $\lambda$ times with new a priori qubit reliabilities\footnote{Here, we prefer the terminology ``qubit reliabilities'' rather than ``input LLRs'' since we modify the actual LLR values.} $ \{ \bRel_{\v}' \}$ and a modified stopping criterion.

For the $k$-th  decoder, the input reliability will be set to $\bRel_{\v}' = 0$ for  qubits  $\v \in \Nbh(c_k)$, considered unreliable, and $\bRel_{\v}' = \bRel_{\v}$ for the rest of the qubits. 
Putting the input reliability to 0 can be considered as an erasure in the MP decoder~\cite{ErasureLDPC}, ensuring that 
these qubits are deprived of any a priori information that may interfere in the decoding attempt of the more reliable qubits.
We further define $\displaystyle \etak = \cup_{\v\in \Nbh (c_k)} \mathcal{N}(\v) $, the set of checks that share a neighbor qubit-node with $c_k$ (note that $c_k \in \etak$). This allows us to define $\bs_{|\overline{\etak}}$ 
the \emph{partial syndrome} vector containing only the checks that have no neighbor qubit-node in $\Nbh (c_k)$, and $\bs_{|{\etak}}$ the \emph{residual syndrome}. 
We then run a MP decoder with a modified stopping criterion that only tries to match the partial syndrome $\bs_{|\overline{\etak}}$. In Algorithm~1, we denote this decoder by $\text{MP}^\star(\mathbf{H}, \bs, \{\mathbf{\bRel}_q'\}, \overline{\etak})$. We emphasize that MP$^*$ applies exactly the same decoding rules on the same Tanner graph as MP, except that MP$^*$ is initialized with qubit reliabilities $\{\mathbf{\bRel}_q'\}$, and it stops when the partial syndrome $\bs_{|\overline{\etak}}$ is satisfied (no matter whether the residual syndrome $\bs_{|{\etak}}$ is satisfied or not). If the MP$^*$ succeeds in matching the partial syndrome, the decoder then attempts to match the residual syndrome by brute-forcing
the error pattern on  $\Nbh (c_k)$.
Note that sometimes the MP$^*$ can actually match the full syndrome, in which case no brute-forcing is needed (will be discussed in more detail later). In Algorithm~1, $\mathbf{H}_{|\overline{\etak}}$ denotes the submatrix of $\mathbf{H}$ whose rows correspond to check-nodes $c\not\in \etak$. Consequently, $\mathbf{H}_{|\overline{\etak}}(c,q) = 0$ for any $q\in\mathcal{N}(c_k)$, and thus $\mathbf{H}_{|\overline{\etak}} \cdot \hat{\mathbf{e}}$ only depends on $\hat{\mathbf{e}}_{|\overline{\mathcal{N}(c_k)}}$ (which explains the slight abuse of notation $\mathbf{H}_{|\overline{\etak}} \cdot \hat{\mathbf{e}}_{|\overline{\mathcal{N}(c_k)}}$). Likewise, $\mathbf{H}_{|\etak}$ denotes the submatrix of $\mathbf{H}$ whose rows correspond to check-nodes $c\in \etak$. If $\hat{\mathbf{e}}_{|\overline{\mathcal{N}(c_k)}}$ matches the partial syndrome, we keep its value and bruteforce $\hat{\mathbf{e}}_{|\mathcal{N}(c_k)}$ to match also the residual syndrome.

\medskip
The intuition behind the post-processing is that the presence of quantum trapping sets \cite{QuantumTrappingSets} in the Tanner graph causes the a posteriori reliability values of trapped qubit-nodes to oscillate. This prevents the decoder from converging, regardless of the number of decoding iterations (for oscillating trapping sets see also~\cite{OscillationTrap}). 
Taking into account the oscillation effect, it is reasonable to think that the messages associated with the untrapped qubits will grow with each iteration while the trapped ones will keep relatively low reliability. This effect will help to identify possible trapped qubits. To this end, we define a reliability metric on checks to decide (the support of) which checks should be \emph{erased}.
%
A natural approach to define such a reliability metric is to consider the reliability (\emph{i.e.}, absolute value) of either incoming messages $\{\mu_{\v\rightarrow c} \mid \v\in \mathcal{N}(c)\}$ or outgoing messages $\{\mu_{c\rightarrow \v} \mid \v\in \mathcal{N}(c)\}$. However, for MS-based decoders, the absolute value of outgoing messages is equal to either the first or the second minimum of the absolute values of incoming messages, denoted by  $\min_{\v \in \Nbh (c)} |\mu_{\v\rightarrow c}|$ and $\minsec_{\,\v \in \Nbh (c)} |\mu_{\v\rightarrow c}|$, respectively. This motivates the reliability metric\footnote{While different variations of this metric are possible (\emph{e.g.}, the sum of the absolute values of all incoming messages)  we have not observed any significant difference in terms of error correction performance. Also,  similar reliability metrics can be obtained for other MP decoding algorithms, \emph{e.g.}, sum-product.} $\delta_c$ considered in Algorithm~1.
%
The cost of computing this metric is nearly none, as the two minima are already computed by the MS decoder. 
Also, this metric is computed for all the check-nodes, based on the $\{\mu_{\v\rightarrow c}\}$ messages at some specific (predetermined) iteration of the MS decoder (as discussed below). 
This allows the sorting of all the checks according to the proposed metric, after which the post-processing can be applied to the $\lambda$ most unreliable checks.


\begin{algorithm}[!t]
\DontPrintSemicolon
\SetAlgoNoLine

\BlankLine

\noindent  $\hat{\mathbf{e}} \leftarrow \text{MP}(\mathbf{H}, \bs, \{\gamma_q\})$

\smallskip\noindent  $\mbox{\textbf{if} } (\mathbf{H} \cdot \hat{\mathbf{e}}  =\bs)  \mbox{ \textbf{then}  } $

\noindent \quad \quad \textbf{return} $\hat{\mathbf{e}} $

\noindent  \mbox{\textbf{else} }

\noindent  \quad \quad  Compute the check reliability values:

\noindent \quad \quad 
$\displaystyle \delta_{c}= \min_{\v \in \Nbh (c)} |\mu_{\v\rightarrow c}| + \minsec_{\v \in \Nbh (c)} |\mu_{\v\rightarrow c}|, \ \forall c \in \mathcal{C}$

\smallskip\noindent \quad \quad   Sort checks in increasing order of reliability, \\
\noindent \quad \quad   Extract $\{ c_k \}_{k\in [\lambda]}$ the least reliable checks.

\smallskip\noindent \quad \quad   {\bf for} $k$ in $1,\dots,\lambda$ {\bf do}


\noindent \quad \quad \quad \quad   $\forall \v \in \mathcal{\V}$, set $\bRel_{\v}' =
\begin{cases}
    0, & \text{if $\v \in \mathcal{N}(c_k)$},\\
    \bRel_\v, & \text{otherwise}.
  \end{cases}$

\smallskip\noindent \quad \quad \quad \quad   Determine $\displaystyle \etak = \cup_{\v\in \mathcal{N}(c_k)} \mathcal{N}(\v) $



\medskip\noindent \quad \quad \quad \quad  $ \hat{\mathbf{e}} \leftarrow \text{MP}^\star(\mathbf{H}, \bs, \{\mathbf{\bRel}_q'\}, \overline{\etak})$

\medskip\noindent \quad \quad \quad \quad  \textbf{if} $ (\mathbf{H}_{|\overline{\etak}} \cdot \hat{\mathbf{e}}_{|\overline{\mathcal{N}(c_k)}}  \neq \bs_{|\overline{\etak}})  \mbox{ \textbf{then}  } $



\smallskip\noindent \quad\quad\quad\quad\quad\quad \textbf{continue }

\noindent \quad \quad \quad \quad  \mbox{\textbf{else} }

\noindent \quad \quad \quad \quad \quad \quad Try to solve 
 $\mathbf{H}_{|\etak} \cdot \hat{\mathbf{e}} =  \bs_{|\etak}$, while 
 
\noindent \quad \quad \quad \quad \quad \quad 
 keeping $\hat{\mathbf{e}}_{|\overline{\mathcal{N}(c_k)}}$ unchanged, and 
 
\noindent \quad \quad \quad \quad \quad \quad  bruteforcing  $\hat{\mathbf{e}}_{|\mathcal{N}(c_k)}$

\smallskip \noindent \quad \quad \quad \quad \quad \quad  \textbf{if}  successful \textbf{then}\\
\noindent \quad \quad \quad \quad \quad \quad \quad \textbf{return }$\hat{\mathbf{e}} $



\smallskip\noindent  $\mbox{\textbf{return} decoding failure}$

\caption{Generic Check-Agnosia Decoder}
\label{alg:bruteforce}
\end{algorithm}

\smallskip
To reduce the overall latency (initial MP decoding and post-processing), one may compute the check reliability values $\delta_c$ at an early iteration, \emph{i.e.}, before the initial MP reaches the maximum number of decoding iterations. 
%
This allows the post-processing to start running in parallel before the initial MP has ended. If the initial MP succeeds later on, the post-processing will stop and the decoder will output the error found by the initial decoder. However, if the initial MP decoder fails, the post-processing will have already started, reducing the total latency. 
As it will be shown in Section~\ref{sec:result}, the error correction performance obtained by determining the list of least reliable checks using the soft information from either the last or an early iteration is almost the same, but the speedup is considerably higher in the latter case.
Moreover, 
the reliability metric computed after a few iterations may be more accurate
 than the one computed at the last iteration, as the oscillation effects (also combined with saturation effects of the finite precision arithmetic)  might alter quite considerably the accuracy of the reliability metric computed after a large number of iterations.

\medskip
We discuss now the brute-forcing of $\hat{\mathbf{e}}_{|\mathcal{N}(c_k)}$ in Algorithm~1. 
To solve the system $\mathbf{H}_{|\etak} \cdot \hat{\mathbf{e}} =  \bs_{|\etak}$ there are several possible methods, including Gaussian elimination. However, since the system to solve is small,  brute-forcing, \emph{i.e.},  trying all the possible combinations, hopefully finding one that satisfies the system\footnote{Note that there is not guaranteed that the system has a solution, as such, the algorithm can fail at this step.}, is a more efficient solution for hardware implementation.
%
Moreover, it is not too difficult to see that 
the brute force approach can be simplified by taking into account the local structure of the code, eliminating a lot of computation.
For instance, a check-node $c\in \etak\setminus \{c_k\}$ that has exactly one qubit-node in common with $c_k$, uniquely determines the value of that qubit.  

\subsection{Check-Agnosia Decoder Without System Solver}
\label{subsec:CA-without-bruteforcing}
One alternative to determine $\hat{\mathbf{e}}_{|\Nbh (c_k)}$, described in Algorithm~2, is to use a regular MP decoder that stops only if the full syndrome is matched. Precisely, the $\text{MP}^\star(\mathbf{H}, \bs, \{\mathbf{\bRel}_q'\})$ in Algorithm~2 is a regular MP decoder, initialized with qubit reliabilities $\{\mathbf{\bRel}_q'\}$, and which stops when the full syndrome is satisfied. We keep the $\text{MP}^\star$ notation in the post-processing step only to distinguish it from the initial MP decoder (will be needed later on Section~\ref{sec:solution}).  To justify Algorithm~2, let us consider the case when the graph induced by any subset $\mathcal{S} \subseteq \mathcal{N}(c_k)$ contains at least a check-node of degree one. 
%
Then, assuming the MP decoder has converged on $\hat{\mathbf{e}}_{|\overline{\Nbh (c_k)}}$, it will converge on the remaining $\hat{\mathbf{e}}_{|\Nbh (c_k)}$  at the cost of a few more iterations. The above condition is the same as requiring  $\Nbh (c_k)$  contains no stopping subset\footnote{A set of qubit-nodes is said to be a stopping set, if the induced subgraph contains no check-nodes of degree 1. If the qubit-nodes in a stopping set are erased, they can get no information during the  MP decoding, that is, incoming and outgoing messages to and from these qubit-nodes remain equal to zero during the entire iterative decoding process.}, and running the MP for a few more iterations amounts to running a peeling decoding~\cite{luby2001efficient} on the erased qubits. For instance, if the Tanner graph contains no cycles of length four, then $\Nbh (c_k)$
satisfies the no-stopping subset condition, and one extra iteration is enough to determine  $\hat{\mathbf{e}}_{|\Nbh (c_k)}$. The no-stopping subset condition may also be satisfied for graphs containing cycles of length four, but in such a case more than one extra iteration may be needed. 

For a given Tanner graph the above no-stopping subset condition can easily be verified, and then we may use Algorithm~2 instead of Algorithm~1 (numerical simulations also confirmed that those two approaches give similar performance). For the simulation results shown later in this paper (Section~\ref{sec:result}), we always use Algorithm~2. 

\smallskip The presumably only meaningful case in which the no-stopping subset condition is not verified is when the code is auto-dual (\emph{i.e.}, $H_x = H_z$), since in such a case  $\hat{\mathbf{e}}_{|\Nbh (c_k)}$ is the support of a codeword, hence a stopping set. It is worth noticing that for auto-dual codes, the check-agnosia (Algorithm~1) and stabilizer-inactivation~\cite{StabilizerInactivationQLDPC} decoders are the same, up to the reliability metric used to select the $\lambda$ least reliable check-nodes. However, for codes that are not auto-dual, the check-agnosia decoder, implemented as in Algorithm~2, presents several advantages, including the use of a simpler, hardware-friendly check-node reliability metric (and not requiring the use of the dual matrix), as well as the fact that it relies solely on MP decoding, eliminating the need of brute-forcing or other system solving methods. 

\begin{algorithm}[!t]
\DontPrintSemicolon
\SetAlgoNoLine
\BlankLine

\noindent  $\hat{\mathbf{e}} \leftarrow \text{MP}(\mathbf{H}, \bs, \{\gamma_q\})$

\smallskip\noindent  $\mbox{\textbf{if} } (\mathbf{H} \cdot \hat{\mathbf{e}}  =\bs)  \mbox{ \textbf{then}  } $

\noindent \quad \quad \textbf{return} $\hat{\mathbf{e}} $

\noindent  \mbox{\textbf{else} }

\noindent  \quad \quad  Compute the check reliability values:

\noindent \quad \quad 
$\displaystyle \delta_{c}= \min_{\v \in \Nbh (c)} |\mu_{\v\rightarrow c}| + \minsec_{\v \in \Nbh (c)} |\mu_{\v\rightarrow c}|, \ \forall c \in \mathcal{C}$

\smallskip\noindent \quad \quad   Sort checks in increasing order of reliability, \\
\noindent \quad \quad   Extract $\{ c_k \}_{k\in [\lambda]}$ the least reliable checks.

\smallskip\noindent \quad \quad   {\bf for} $k$ in $1,\dots,\lambda$ {\bf do}

\noindent \quad \quad \quad \quad   $\forall \v \in \mathcal{\V}$, set $\bRel_{\v}' =
\begin{cases}
    0, & \text{if $\v \in \mathcal{N}(c_k)$},\\
    \bRel_\v, & \text{otherwise}.
  \end{cases}$

\medskip\noindent \quad \quad \quad \quad  $ \hat{\mathbf{e}} \leftarrow \text{MP}^\star(\mathbf{H}, \bs, \{\mathbf{\bRel}_q'\})$

\medskip\noindent \quad \quad \quad \quad  \textbf{if} $ (\mathbf{H} \cdot \hat{\mathbf{e}}  = \bs)  \mbox{ \textbf{then}  } $

\noindent \quad\quad\quad\quad\quad \textbf{return} $\hat{\mathbf{e}} $


 
 


\smallskip\noindent  $\mbox{\textbf{return} decoding failure}$

\caption{\parbox[t]{48mm}{Check-Agnosia Decoder Without System Solver}}
\label{alg:no-bruteforce}
\end{algorithm}

A final remark is that all MP and $\text{MP}^\star$ decoders can implement a flooded or a layered schedule, as discussed in Section~\ref{sec:algos}, to cope with the hardware constraints. 
\section{Hardware Architectures}\label{sec:solution}

This section aims to analyze the impact of the post-processing algorithm on the hardware implementation, considering architectures with different schedules and varying degrees of parallelism. We carry out a detailed analysis of different corner cases,  providing latency and power bounds to assist future hardware decoder designers.

\subsection{MP Decoder Architecture}
\label{subsec:mp-architecture}
We consider first a single MP decoder, without any post-processing. To implement the MP decoder in hardware \footnote{Serial schedule is not considered due to its extremely large latency, not suitable for real-time implementations.}, one can use a fully parallel architecture, implementing a flooded schedule, referred to as flooded decoder, or a partly parallel architecture, implementing a layered schedule, referred to as layered decoder. 

We will make standard assumptions\footnote{The hardware implementation reported later on Section~\ref{subsec:implem-results} is consistent with the assumptions made here.} regarding the two above architectures~\cite{boutillon2014hardware}. For the flooded decoder, the Tanner graph is instantiated in hardware, where messages are exchanged through wires between processing units, corresponding to qubit- and check nodes. Each decoding iteration is performed in two clock cycles, with one clock cycle for qubit-node messages and a posteriori LLRs, and a second one for check-node messages.
Thus, the worst case (maximum) latency of the flooded decoder is equal to $(1 + 2I_{F})/f_F$ (s), where we count one clock-cycle for data loading, $I_F$ is the maximum number of decoding iterations of the flooded decoder, and $f_F$ is the clock frequency. 

For the layered decoder, the number of processing units instantiated in hardware is given the size of the largest layer\footnote{Usually all layers have the same size, although this condition is more difficult to satisfy for qLDPC codes~\cite{23Layered}.},  messages are exchanged through shared memory, and each processing unit is reused $\eta_L$ times for each decoding iteration, where $\eta_L$ denotes the number of layers per iteration.   The worst case latency of the layered decoder is equal to $(1 + \eta_L I_{L})/f_L$ (s), where we count again one clock-cycle for data loading, $I_L$ is the maximum number of decoding iterations of the layered decoder, and $f_L$ is the clock frequency.


\smallskip Two observations are in place here. First, the flooded architecture may lead to a large number of connections among processing units, causing routing congestion in case of large codes. Due to the large interconnect network, the operating clock frequency of the flooded architecture ($f_F$) is usually smaller than twice\footnote{Note that in the flooded architecture, qubit and check-node messages are computed in two different clock cycles, by different processing units, while in the layered architecture they are computed in the same clock cycle, by a processing unit that merges the qubit and check node processing.} that of the layered architecture ($f_L$). Second, as discussed in Section~\ref{sec:algos}, the layered schedule propagates information about twice faster than the flooded one, thus the maximum number of iterations of the layered architecture ($I_L$) is usually smaller than that of the flooded architecture ($I_F$). Overall, this can make the layered architecture comparably fast to the flooded one, despite the fact that it employs a reduced degree of parallelism (of course, the number of layers per iteration has to be sufficiently small). 

\smallskip 
Finally, one possible approach to further increase the clock frequency of the layered decoder is to pipeline the design (\emph{i.e.}, perform each layer in a number of pipelined clock cycles). However, this may lead to  delayed
message write-backs in memories, and thus, to pipeline related
hazards~\cite{boncalo2018layered}.  Solving such hazards (without relying on pipeline stalls, introducing extra latency) can be done for classical LDPC codes at the code construction stage~\cite{boncalo2018code}. However such solutions are not generic (need a specific code construction) and may not apply to qLDPC codes. Therefore, to keep the analysis as generic as possible, we do not consider pipelined designs in this work.

\subsection{Post-Processing Elements}
\label{subsec:postproc-elements}
For the check-agnosia scheme, the first step after the MP decoder is the computation of the check reliability values, as outlined in Algorithms 1 and 2. The metric used to calculate the check reliability, denoted as $\delta_c$, involves adding the two least reliable messages. These values are computed during the tree finder process employed to calculate check-node messages in the min-sum decoder. Thus, the only additional hardware required is an adder per check-node to compute $\delta_c$. These values are updated on-the-fly during each iteration, eliminating the need for extra clock cycles after the MP decoder. As described earlier in Section~\ref{sec:arch}, one does not have to wait until the end of the initial MP decoder to start the post-processing (the impact of utilizing the $\delta_c$ information from early iterations will be evaluated in Section~\ref{sec:result}).  
In the proposed architecture, the $\delta_c$ values can be stored in the registers of the sorting unit (see below) before the first MP decoder completes, without any additional hardware. This allows absorbing some additional latency and initiating the post-processing $\text{MP}^\star$ decoders before the initial MP decoder completes.

After the $\delta_c$ values are available, a sort of the checks in order of reliability is computed. To sort the checks in order of increasing reliability |$\mathcal{C}$|-1 comparators are required to implement a tree structure, which should be pipelined to avoid increasing the critical path of the decoder. The number of clock cycles needed to obtain the complete sorted list is $\left\lceil\lambda/2\right\rceil \times \left\lceil\text{log}_2 |\mathcal{C}|\right\rceil$. 

\subsection{Overall Check-Agnosia Architecture}
\label{subsec:CA-architecture}
\begin{figure}[!t]
\centering
\includegraphics[width=1.0\linewidth]{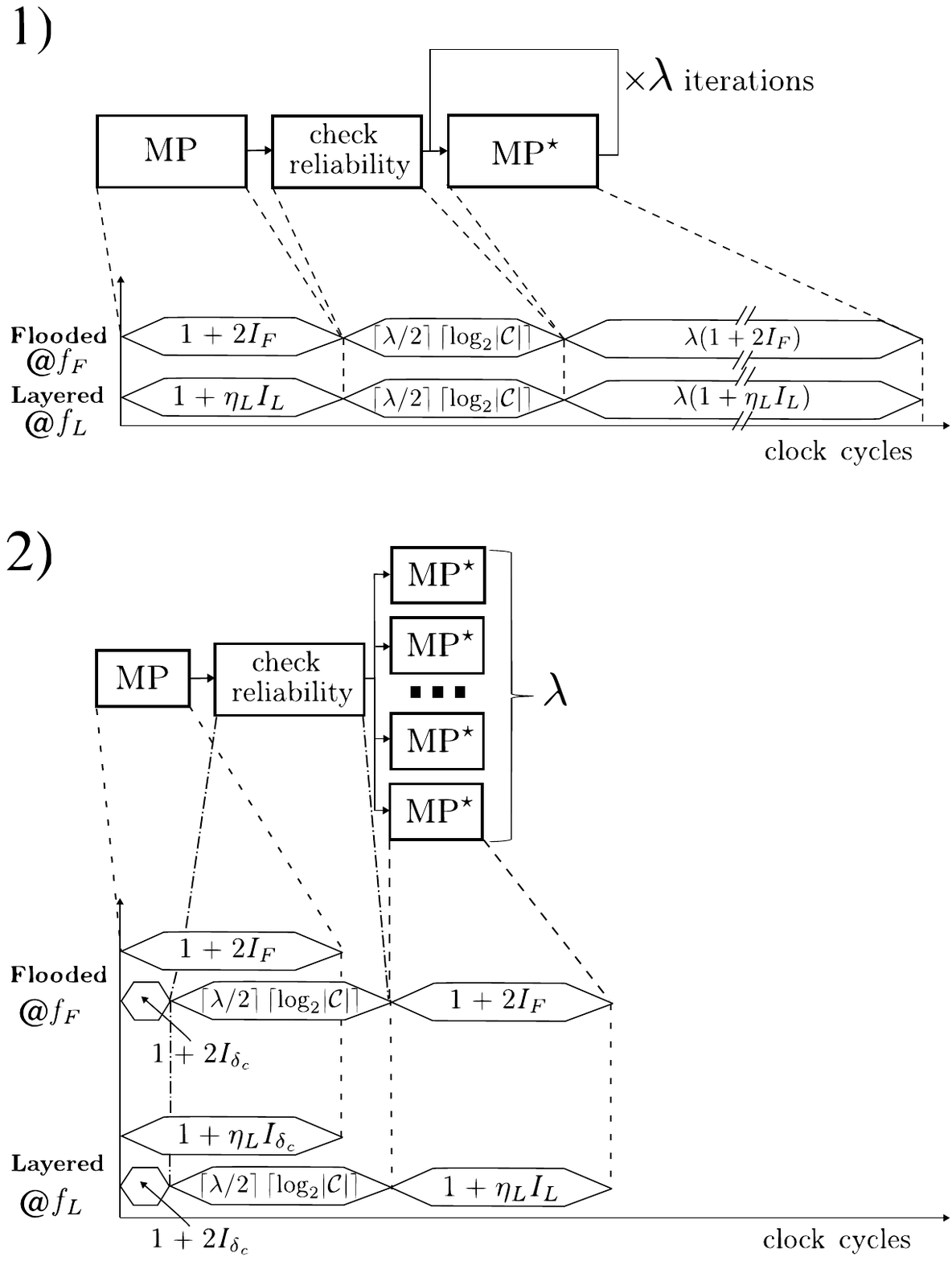}
\caption{Comparison of different architectures for the check-agnosia decoder. The clock cycle diagram is included for the different proposals (Warning: drawing is not to scale). In case 1), MP and MP$^\star$ use the same hardware. 
}
\label{fig:architectures}
\end{figure}

In this section, we detail the check-agnosia architecture corresponding to Algorithm~2 (that relies on MP decoding only, without brute-forcing). After the list of $\lambda$ least reliable checks is obtained, the $\lambda$ $\text{MP}^\star$ decoders are performed. Depending on the time constraints and/or power budget, we may consider two different approaches, illustrated in Figure~\ref{fig:architectures}. 
 
 The first approach consists of performing the $\lambda$ $\text{MP}^\star$ decoders sequentially, reusing the same hardware as the one used for MP. Only |$\mathcal{\V}$| extra multiplexors are required to choose between $ \bRel_{q}'= \bRel_{q}$ or $ \bRel_{q}'= 0$, and $|\mathcal{C}|$ extra multiplexors are required to decide which syndromes belong to $\bs_{|\etak}$, depending on the check $c_k$. This approach of reusing hardware yields higher latency, but maybe interesting for a quantum computer with time constraints close to microseconds, \emph{e.g.}, based on trapped ion technology~\cite{ChallengesZooImplementQLDPC}. 

 For the second approach, the $\lambda$ $\text{MP}^\star$ decoders are performed in parallel, by using dedicated hardware. Moreover, the $\lambda$ $\text{MP}^\star$ decoders may start before the initial MP completes, using check-reliability values computed at an early iteration, that we will denote in the sequel by $I_{\delta_c}$.  This approach may be interesting for quantum technologies with more restrictive latency constraints,  but having in mind that power can be also a limitation, as happens with superconducting qubits in which the decoder needs to reduce its power budget when it operates close to the quantum chip at cryogenic temperatures. 

\smallskip To illustrate the degree of complexity in hardware implementations and measure the gap between the proposed solutions to latency/power constraints, we analyze below the Pareto designs for the two approaches above, where the MP decoder uses either a flooded or a layered schedule. We provide the worst-case latency (simply referred to as latency), as well as the power consumption as a function of the nominal power consumption of the MP decoder, denoted by $P_F$ or $P_L$, with a subscript indicating the flooded or layered architecture (we may reasonably assume that MP and MP$^*$ yield the same power consumption). For the latency value, we take into account the latency induced by sorting the check nodes according to their reliability (Section~\ref{subsec:postproc-elements}). The corresponding power consumption is not accounted for, we will assume it is negligible with respect to the power consumption of the MP decoder. 


\medskip\noindent{\bf Flooded MP/MP$^*$ decoders:}
\begin{enumerate}
    \item Hardware reuse (sequential post-processing)
    
    MP flooded decoder + check reliability unit + one $\text{MP}^\star$ flooded decoder running $\lambda$ rounds:
    \begin{itemize}
        \item Latency: $\big[(1+2 I_F) \ + \  \left\lceil\lambda/2\right\rceil \left\lceil\text{log}_2 |\mathcal{C}|\right\rceil \ + 
        \lambda (1+2 I_F)\big] / f_F$
        \item Power: $P_F$ 
    \end{itemize}
    
    \item Dedicated hardware  (parallel post-processing) 
    
    MP flooded decoder + check reliability unit starting after iteration $I_{\delta_c}$ + $\lambda$ $\text{MP}^\star$ flooded decoders running in parallel
    \begin{itemize}
        \item Latency: $\big[(1+2 I_{\delta_c}) \ + \  \left\lceil\lambda/2\right\rceil \left\lceil\text{log}_2 |\mathcal{C}|\right\rceil \ + 
        (1+2 I_F)\big] / f_F$
        \item Power: $(\lambda+1) P_F$
    \end{itemize}
\end{enumerate}

\medskip\noindent{\bf Layered MP/MP$^*$ decoders:}
\begin{enumerate}
    \item Hardware reuse (sequential post-processing)
    
    MP layered decoder + check reliability unit + one $\text{MP}^\star$ layered decoder running $\lambda$ rounds:
    \begin{itemize}
        \item Latency: $\big[(1+\eta_L I_L) \ + \  \left\lceil\lambda/2\right\rceil \left\lceil\text{log}_2 |\mathcal{C}|\right\rceil \ + 
        \lambda (1+\eta_L I_L)\big] / f_L$
        \item Power: $P_L$ 
    \end{itemize}
    
    \item Dedicated hardware  (parallel post-processing) 
    
    MP layered decoder + check reliability unit starting after iteration $I_{\delta_c}$ + $\lambda$ $\text{MP}^\star$ layered decoders running in parallel
    \begin{itemize}
        \item Latency: $\big[(1+\eta_L I_{\delta_c}) \ + \  \left\lceil\lambda/2\right\rceil \left\lceil\text{log}_2 |\mathcal{C}|\right\rceil \ + 
        (1+\eta_L I_L)\big] / f_L$ 
        \item Power: $(\lambda+1) P_L$
    \end{itemize}
\end{enumerate}

 \subsection{Implementation Results}
\label{subsec:implem-results}

To illustrate the analysis from the previous section, we have implemented both flooded and layered NMS decoders on a Xilinx FPGA xcv095 board, for the B1[[882, 24]] code from~\cite{DegenerateQLDPC}. The implemented decoders use finite precision arithmetic, with exchanged messages quantized on $6$ bits, and a posteriori LLR values quantized on $8$ bits. The parity-check matrix (for both $X$ and $Z$ errors) is of size $441\times 882$ (check-nodes $\times$ qubit-nodes) and has no four-cycles (thus, it satisfies the no-stopping subset condition, and we may safely apply Algorithm~2). 


The flooded NMS / NMS$^*$ decoders achieve a maximum operating frequency $f_F= 100$\,MHz (corresponding to a critical path of 10\,ns), with 62\% of the hardware resources of the device utilized, and a total power consumption  $P_F=5.5$\,W.

To implement the layered decoders, we use the $2$-covering approach from~\cite{23Layered}, where $7$ overlapping layers are used to cover $2$ iterations, yielding a fractional number of layers per iteration $\eta_L = 3.5$.  The layered NMS / NMS$^*$ decoders achieve a maximum operating frequency  $F_L= 80$\,MHz (corresponding to a critical path of 12.5\,ns), where about  25\% is due to the logic depth of the operations and 75\% is due to the routing limitations of the FPGA device. The decoder uses only 13\% of the hardware resources of the device, and the total power consumption is around $P_L=2.03$\,W.

\smallskip We consider a maximum number of decoding iterations $I_F = 30$ for the flooded decoders, and $I_L = 15$ for the layered decoders (due to faster convergence). For the post-processing step, we consider a list of $\lambda = 10$ least reliable checks (these parameters will be evaluated from the error correction perspective in Section~\ref{sec:result}). 
Latency and power consumption values are summarized in Table~\ref{tab:implem-results}, for the Pareto designs considered in the previous section. Note that we consider two cases for the dedicated hardware scenario, in which the iteration $I_{\delta_c}$ (used to compute the check-node reliability values) is chosen to be either the last or the third iteration of the NMS decoder. 

\begin{table}[!t]
\caption{Latency ($L$) and power consumption ($P$) values for the Pareto designs in Section~\ref{subsec:CA-architecture} ($I_\text{max}$ is the table stands for $I_F$ for flooded architectures, or for $I_L$ for layered ones). }
    \label{tab:implem-results}
    \centering
    \begin{tabular}{|c|c|c|}
    \cline{2-3}
     \multicolumn{1}{c|}{}      &  Flooded  & Layered \\
    \hline
 HW reuse  &  $L = 7.2\,\mu$s &  $L = 7.9\,\mu$s\\
           &  $P = 5.5$\,W &  $P = 2.03$\,W \\
 \hline
 Dedicated HW & $L = 1.7\,\mu$s & $L = 1.9\,\mu$s\\
    $I_{\delta_c} = I_{\text{max}}$    &  
     $P = 60.5$\,W &  $P = 22.3$\,W \\
    \hline
    Dedicated HW &  $L = 1.1\,\mu$s &  $L = 1.4\,\mu$s\\
    $I_{\delta_c} = 3$    & $P = 60.5$\,W &  $P = 22.3$\,W \\
    \hline
    \end{tabular}
    
\end{table}



\smallskip It can be observed that the layered architecture achieves latency values close to the flooded one, despite the fact it employs a degree of parallelism $3.5$ times lower, while considerably reducing the power consumption. It is also worth noticing that the part of the latency due to the sorting unit is 0.45\,$\mu$s for the flooded architectures, and 0.56\,$\mu$s for the layered ones. To reduce the latency of the sorting unit further optimizations are possible (\emph{i.e.}, carefully balancing the pipeline stages of the sorting unit by taking into account the maximum critical path latency of the MP decoder, splitting the sorting unit into layers in case of a layered schedule, or using a different clock domain for the sorting unit), which are however behind the scope of this work. We mention that the maximum frequency that can be reached for the sorting unit (implemented alone) is 230\,MHz, which gives a lower bound on the achievable latency of 0.2\,$\mu$s\footnote{At an operating frequency of 230\,MHz, the power consumption for the sorting unit is 0.66\,W,   versus 0.26\,W at at 100\,MHz.}. 

\smallskip Moreover, as will be shown in Section~\ref{sec:result}, because of the highly degenerate structure of the codes, the layered schedule provides better error correction performance than the flooded one, even if the number of decoding iterations of the latter exceeds significantly the number of decoding iterations of the former (in fact, to get a flooded decoder that approaches the layered decoder, albeit not closely, one would have to go for at least 60 iterations, see Section~\ref{sec:result}). One last advantage of the layered architecture, reported in~\cite{23Layered}, is that the logical error rate can be considerably improved by processing layers in random order at each iteration. Such a random layer order can be implemented at a very low cost, as it only requires modifying the ROM memory that stores the layers' control sequence and including a deeper memory with a pseudo-random sequence of layers.



\smallskip From the results presented before, it can be concluded that timing constraints can be in the range of the requirements reported in \cite{HWSurveyQEC} for transmons and ion trap technology, between microseconds and milliseconds. However, these implementations do not meet the highly restrictive conditions of superconducting qubits in both time and power which are around 400\,ns and 1W, see \cite{PowerLimit}. The difference compared to the fastest solution in Table~\ref{tab:implem-results} exceeds 3 times the time budget and it is more than one order of magnitude far in terms of power consumption. For these scenarios, it is important to remark that other approaches to implementation like ASICs or more advanced FPGA devices based on 16nm CMOS process or below (note that the xcvu095 belongs to the previous generation of 20nm) need to be explored in future work. Moreover, exploiting a ping-pong architecture that takes benefit of the pipeline registers to reduce the number of $\text{MP}^\star$ decoders to half for the parallel implementation of flooded schedule can be a good proposal to reduce power consumption to almost half. 


\smallskip Extrapolating from state-of-the-art ASIC implementations of classical LDPC decoders, a clock frequency of 151\,MHz is reported in~\cite{nguyenly2017analysis} for a 65-nm CMOS ASIC implementation of a min-sum decoder using the layered architecture described in Section~\ref{subsec:mp-architecture}, for a regular LDPC code with characteristics similar to those of the B1 code investigated here\footnote{In~\cite{nguyenly2017analysis}, the parity check matrix is of size $648 \times 1296$, with column weight 3 and rows weight 6. Each layer consists of 216 checks (referred to as full-layer therein). The parity check matrix of the B1 code is of size $441 \times 882$, with column weight 3 and rows weight 6, and each layer consists of 126 checks.}. The operating frequency is expected to further increase for the B1 code, given that both the parity check matrix and the layer size are smaller than that of the LDPC code in~\cite{nguyenly2017analysis}.  For $f_L = 151$\,MHz, the latency of the layered architecture with parallel post-processing (dedicated hardware) is equal to $1\,\mu$s if $I_{\delta_c} = I_L$ (last iteration), and $0.73\,\mu$s if $I_{\delta_c} = 3$. Since the operating frequency increases with decreasing technology node, and assuming an inverse-linear frequency scaling~\cite{hauser2008mosfet}, we may conclude that a latency constraint around 400\,ns or below can be easily achieved for more advanced technology nodes, \emph{e.g.}, below 22\,nm (today technology scaling is actually much lower).

Finally, we note that all the previous results assume that the check-agnosia decoder is implemented as in Algorithm~2. 
For the codes where Algorithm 2 cannot be applied, the latency will be a little bit worse than what was computed here, due to the brute-forcing step in Algorithm~1.
We also provide in Appendix~\ref{appendix:osd}  arguments for why OSD post-processing (widely used today in the community for decoding of small to medium LDPC codes) is not a viable solution going forward, if trying to cope with the hardware implementation constraints.

\section{Error Correction Performance}\label{sec:result}

\begin{figure*}
\centering
\subfloat[{$B1[[882,24]]$ flooded}]{\includegraphics[width=0.5\linewidth]{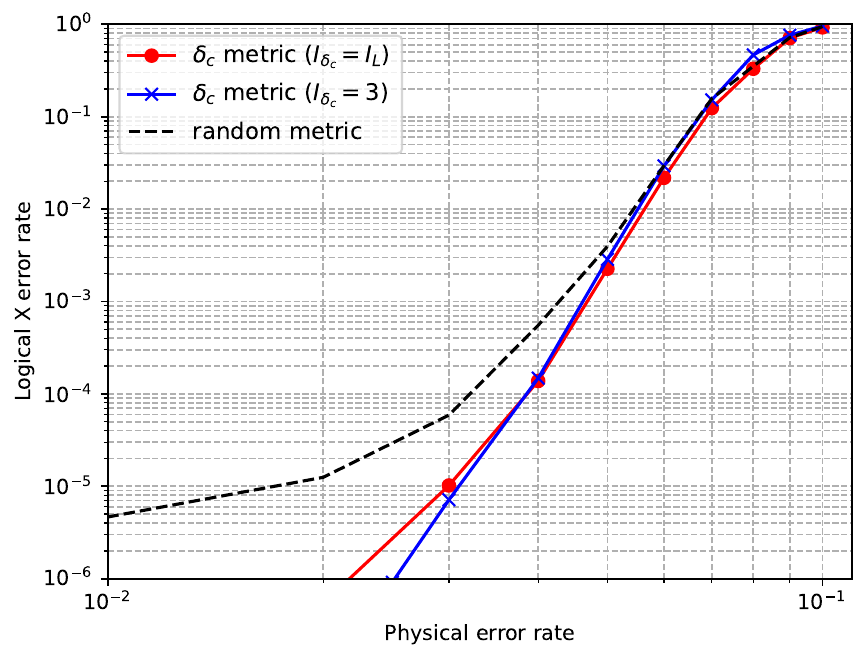}}\hfill%
\subfloat[{$B1[[882,24]]$ layered}]{\includegraphics[width=.5\linewidth]{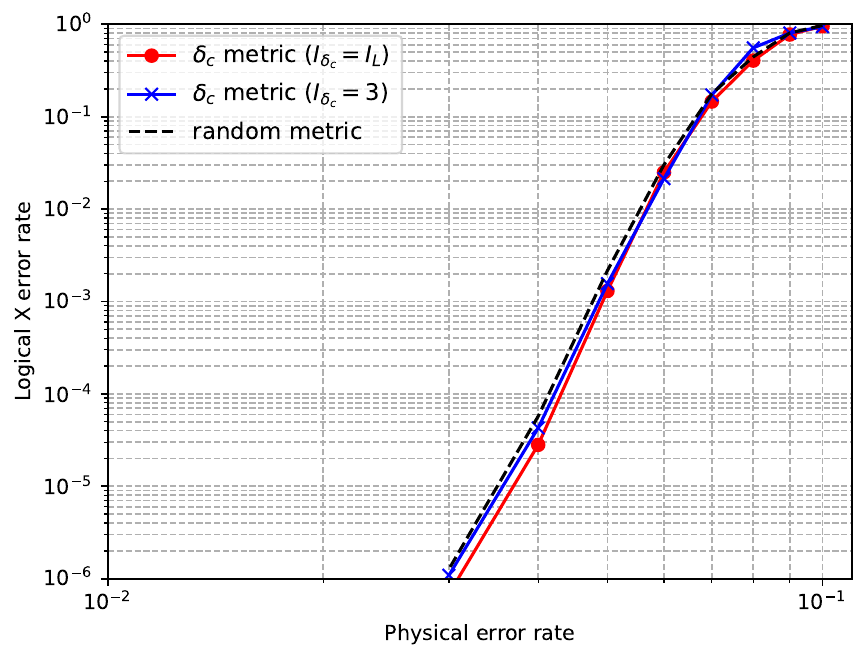}}\\
\caption{Performance of the check-agnosia decoder with different reliability metrics (B1 code)}
\label{fig:ppiter}
\end{figure*} 

\begin{figure*}
\centering
\subfloat[{ $B1[[882,24]]$}]{\includegraphics[width=0.5\linewidth]{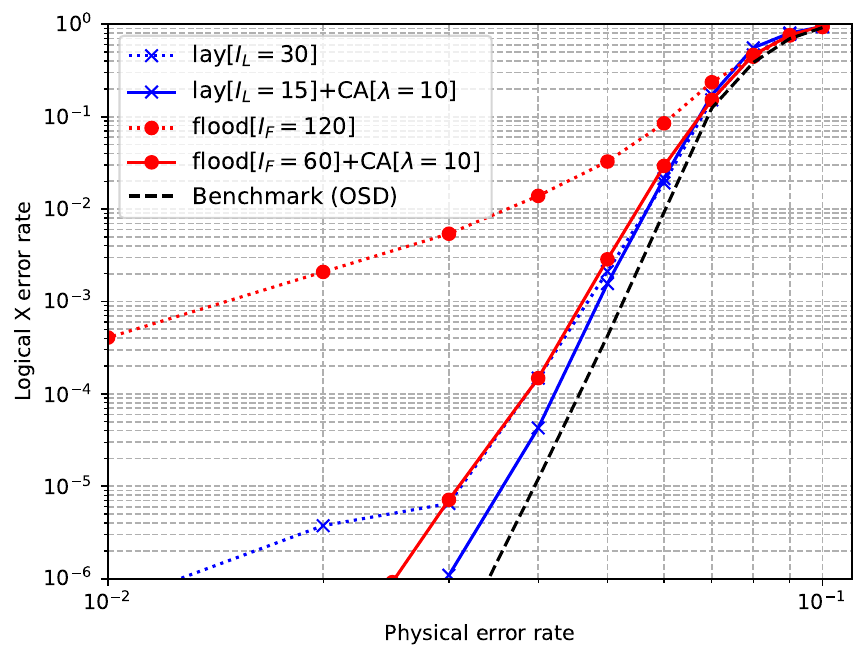}}\hfill%
\subfloat[{$C2[[1922,50,16]]$}]{\includegraphics[width=.5\linewidth]{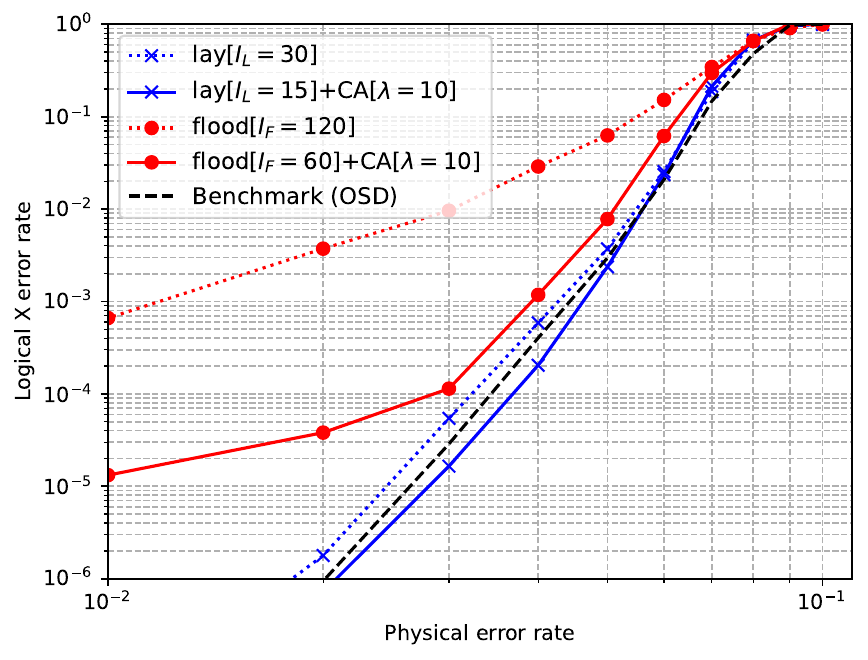}}\\
\caption{Analysis of the check-agnosia post-processing on  codes B1 and C2 ($\delta_c$ metric, with $I_{\delta_c}=3$).
}
\label{fig:num_res}
\end{figure*} 

In this section, we evaluate the error correction performance of the proposed check-agnosia post-processing. The codes used are B1[[882, 24]] and C2[[1922, 50]] from~\cite{DegenerateQLDPC}. For both codes, the no-stopping subset condition from Section~\ref{subsec:CA-without-bruteforcing} is satisfied, hence in the following all simulations are performed using Algorithm~2 (without brute-forcing the system).

As our post-processing is targeted at decoding $X$ and $Z$ errors separately, we use an $X$ noise model, and thus ``physical error rate''  does actually refer to the physical $X$ error rate.

Our numerical simulations are consistent with the parameters used in Section~\ref{subsec:implem-results}. Precisely, we consider a finite-precision NMS decoder, using 6 bits for the exchanged messages, and 8 bits for the a posteriori LLRs. Although in floating point precision the initial (a priori) LLRs of the NMS decoder can be scaled to 1, in finite precision the initial LLR values have a non-negligible impact on convergence. 
In the simulations, we use the following parameters, optimized by extensive search. 
For the flooded decoder we set the initial LLR values to $\text{LLR}_\text{init}=12$ and the NMS scaling factor is set to $s_\text{\sc nms}=0.875$. For the layered decoder we use $\text{LLR}_\text{init}=8$ and NMS scaling factor $s_\text{\sc nms}=0.9375$. Note that scaling factors are a sum of powers of 2 and as such the scaling operation can be implemented efficiently in hardware, using only SHIFT and ADD operations.

For the maximum number of decoding iterations, we use $I_F = 60$ for the flooded decoder, and $I_L=15$ for the layered decoder. The maximum number of decoding iterations for the flooded decoder is the only deviation with respect to the parameters used in Section~\ref{subsec:implem-results} (where $I_F = 30$ was used). In fact, our goal here is to demonstrate the advantage in terms of error correction performance of the layered architecture as compared to the flooded one, even when the latter employs a significantly higher number of decoding iterations.  For the post-processing part, we use $\lambda = 10$.

Whenever the layered decoding is used, we add the random ordering perturbation introduced in \cite{23Layered} that was shown to significantly improve the decoding convergence.

Figure \ref{fig:ppiter} shows the impact of the iteration $I_{\delta_c}$ used to select the checks in the post-processing, all simulations are done on the B1 code. 
For flooded simulations (Figure \ref{fig:ppiter}(a)),  it is actually beneficial to use the 3rd iteration for the metric instead of the last (60th iteration). The most probable explanation is that the relatively high number of iterations combined with the finite precision algorithm makes the metric less reliable after a larger number of iterations. For comparison purposes, we also consider a random metric, corresponding to a random choice of the $\lambda$ checks in the post-processing. 
As it can be seen, the random metric exhibits a bad error floor, validating the metric used in that case.

For layered simulation (Figure \ref{fig:ppiter}(b)), all three curves are close by, since the layered NMS decoder with random layer ordering performs already very well.

In fact, the three metrics yield virtually the same performance of the check-agnosia decoder, but which is better than the layered NMS with 30 iterations and without post-processing in Figure~\ref{fig:num_res}(a). Although the metric is less important in this case, this shows that the perturbation introduced by the post-processing step in the input reliabilities has an impact on the decoding, and that it is better to run multiple decoders in parallel with perturbed inputs and fewer iterations rather than running a single decoder for a long time. 
As a whole, this validates the fact that the post-processing can be done efficiently using the dedicated hardware approach, increasing the post-processing parallelism and improving the latency.

We would also like to make a case for the choice $I_{\delta_c}=3$. This hyperparameter can be optimized to get the best numerical results for a given code. However, the value 3 here was chosen for a different reason. Since both codes have girth 6, choosing $I_{\delta_c}$ to be equal to 3 guarantees that when the aposteriories are extracted, the decoder got access to the information of the biggest neighbourhood of each variable nodes \textit{without loopy information}. 
This ensures that although it is very local, this information is also less noisy than information coming from later rounds.

In Figure~\ref{fig:num_res}, the post-processing is applied to the codes B1 and C2, with both flooded and layered schedules. For a comparison with the state of the art, in both figures, we added a dashed black curve of an optimized NMS-OSD decoder using 100 iterations, floating point NMS with a scaling factor of 0.625~\cite{DegenerateQLDPC}. Keep in mind that this decoder is not at all hardware-friendly,  in terms of complexity, latency and power consumption, and it only serves as a reference. As it can be seen from both simulations, our results are matching closely the performance of the OSD post-processing, concretely showing the effectiveness of our hardware-friendly approach.
In both figures, the results in red are the curves for flooded and in blue for layered. Each time, the dotted curves show the performance of the decoder without post-processing.

On the B1-code for the flooded schedule, the impact of the post-processing is clear, and the check-agnosia flooded decoder exhibits good performance while keeping a latency around $1.7\,\mu$s (taking into account $I_F=60$). For the layered schedule, the use of the post-processing increases the steepness of the waterfall. The check-agnosia layered decoder keeps the latency at around $1.4\,\mu$s. (Latency values above correspond to our FPGA implementation from Section~\ref{subsec:implem-results}.) 

On the C2 Code, the performance gains for flooded are clear even if the post-processing suffers from a relatively high error floor. For layered scheduling, once again check-agnosia achieves better results in the error floor compared to no post-processing, closely matching the NMS-OSD curve.

Further numerical results are provided in Appendix~\ref{appendix:lambda}, where we evaluate the error correction performance of the check-agnosia decoder on the family of T-codes from~\cite{BPOSD}, showing a threshold phenomenon with  
hyperparameter $\lambda=0.02\times |C|$.

\section{Conclusions}\label{sec:conclusion}

This work introduced the check-agnosia algorithm, a new post-processing method improving on the syndrome-inactivation algorithm from a hardware-oriented viewpoint. Interestingly, although in the general case brute-forcing a small linear system may still be needed, for a large class of qLDPC codes the check-agnosia post-processing relies only on MP decoding, eliminating the need for any system solver. 
  The proposed solution is flexible and it allows devising different hardware architectures, in order to meet the latency or the power constraints of the quantum system. The analysis carried out in the document, along with the hardware implementation results for MP decoders (our own implementation on an FPGA board, or results extrapolated from state-of-the-art ASIC implementations), showed that our solution can meet latency constraints of a wide range of quantum technologies, while providing state of the art error-correction performance, with hardware-accurate, finite-precision arithmetic.  
To the best of our knowledge, there is no prior work on the hardware architecture and implementation of a post-processing enhanced MP decoder for qLDPC codes. 
An interesting open question going forward would be to look at space-time decoding and see if the underlying graph structure lends itself well to the use of the check-Agnosia post-processing without system-solving. 

\section{Acknowledgment}

This work is supported by the QuantERA grant EQUIP (French ANR-22-QUA2-0005-01 and Spain MCIN/AEI/10.13039/501100011033), by the Plan France 2030 (ANR-22-PETQ-0006), the grant PCI2022-132922 funded by Agencia Estatal de Investigación, Ministerio de Ciencia e Innovación, Gobierno de España and by the European Union “NextGenerationEU/PRTR”.

\appendix

\section{Hyperparameter $\lambda$ and decoding threshold}
\label{appendix:lambda}

\begin{figure}[!b]
\centering
\includegraphics[width=\linewidth]{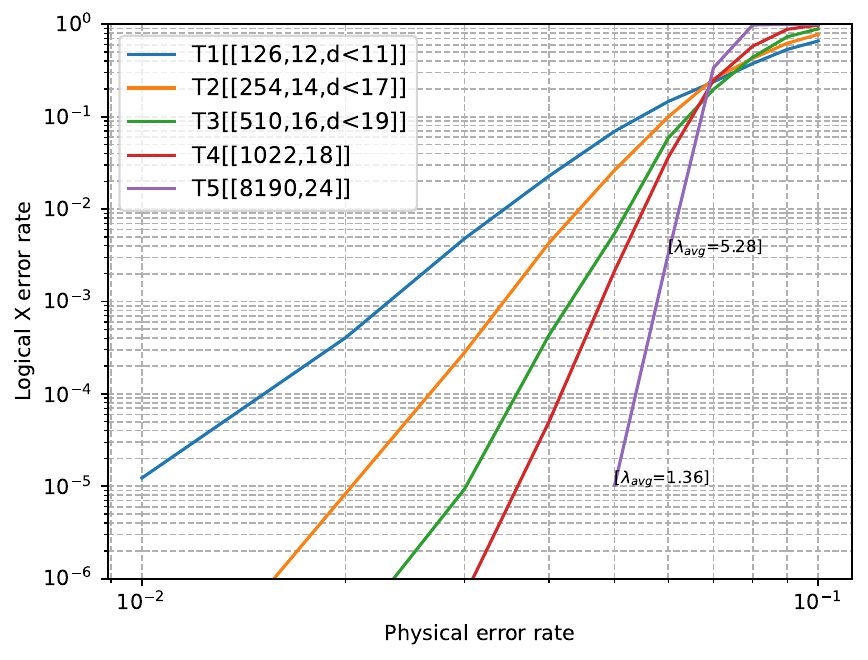}
\caption{ Check-agnosia threshold for the  T codes family from~\cite{BPOSD}. Serial scheduling with random ordering and 15 iterations. Normalized min-sum with scaling factor 0.9375 and finite precision arithmetic, with 6-bit quantization for the input LLRs and exchanged messages, and 8-bit quantization for the a posteriori LLRs.  For the post-processing, check-agnosia is used with $I_{\delta_c} = 3$ and $\lambda=0.02\times|\mathcal{C}|$.}
\label{fig:threshold}
\end{figure} 

In Figure~\ref{fig:threshold}, we include additional numerical results giving a threshold for a constant rate family of LDPC codes, namely the T codes family from~\cite{BPOSD}. Since we are not aware of a simple way to build layers for this family of codes, we ran the simulations using a serial decoder, which is a fair approximation of the numerical results one would get with layered decoding. We use check-agnosia with hyperparameter $\lambda=0.02\times |\mathcal{C}| = 0.01 \times |\mathcal{Q}|$, since for all the codes of the T family,  $|\mathcal{C}| = |\mathcal{Q}|/2$.  The computational complexity of the algorithm hence is $(0.01\times) n^2 \times \log n $, where $n= |\mathcal{Q}|$ is the number of qubits, and this complexity can be spread between time and energy consumption depending on the architecture needs (see Fig~\ref{fig:architectures}). Furthermore, we make a case that the average complexity of the decoder is actually much better than that. On the figure, we also included the average number of inactivations (denoted $\lambda_{avg}$) for physical error rates 0.6 and 0.5, where the average values get very close to one (meaning only one inactivation might usually be necessary). Since this number goes close to one for low error-rates, it means that in practice the cost of the post-processing could only add a constant multiplicative overhead. This lambda average is particularly meaningful in a sequential architecture where we stop the post-processing as soon as the first post-processing converges. In the parallel architecture, it should still be possible to optimize the actual number of parallel runs if we have access to some prior information on the noise level, \emph{e.g.}, by considering a pool of MP$^\star$ decoders that serve for the post-processing of several logical qubits and are dynamically allocated between them.

\section{Latency comparison of CA and OSD}
\label{appendix:osd}

We provide below  
a comparison, in terms of latency, between the check agnosia proposal and the OSD post-processing solution. This comparison is similar to the method presented in \cite{OSD_FPGA}, and is intended to clarify the differences between the two solutions with respect to hardware VLSI implementations.

 We consider a layered MP decoder, with check-agnosia implemented through ``Dedicated hardware'' (that is, the $\lambda$ $\text{MP}^\star$  decoders are executed in parallel). Since the  layered $\text{MP}^\star$ decoders achieve a maximum operating frequency $F_L= 80$\,MHz (corresponding to a critical path of 12.5\,ns), the total latency of the check-agnosia post-processing is $(12.5 \times 3.5 \times 15) = 656.25$\,ns. We have omitted here the latency of the sorting unit, required to sort the check-nodes according to their reliability.

 Considering now the OSD post-processing, we will omit again the latency of the sorting unit, required this time to sort the qubit-nodes according to their reliability. We will actually consider only the latency of the Gaussian elimination step required by OSD post-processing (and omit the latency of any other steps). Refs.~\cite{Rupp2006Parallel, Hochet87} below provide the two main highly parallel architectures known in the literature to perform Gaussian elimination over finite fields. However, in both cases, the number of clock cycles required to perform Gaussian elimination is equal to $(M^2+M)/2$, where $M$ is the number of rows of the parity-check matrix. Thus, the latency of the Gaussian elimination implementation is determined by $T_\text{OSD}=(M^2+M)/2/f_\text{OSD}$, where $f_\text{OSD}$ is the operating frequency. So the frequency required to achieve the same time budget as our proposal is $f_\text{OSD} = (441^2+441)/2/ 656.25\,\text{ns} = 148.5\,\text{GHz}$. Such a frequency is completely unrealistic, and would certainly lead to timing violations in the design (note that it is $148.5/0.08=1856$ times larger than the one of the layered MP decoder). Besides, it would also translate into an extremely large power consumption, which typically increases linearly with the operating frequency.

\bibliographystyle{IEEEtran}
\bibliography{biblo.bib}

\end{document}